\documentclass[prl,superscriptaddress,amsmath,amssymb,floatfix,twocolumn,showpacs,amsfonts,longbibliography,prx]{revtex4-2}
\usepackage{times}
\usepackage[varg]{txfonts}
\usepackage{textcomp}
\usepackage{graphicx}
\usepackage{subfigure}
\usepackage{tabu}
\usepackage{color}
\usepackage[plainpages=false,pdfpagelabels,colorlinks=true,linkcolor=red,urlcolor=blue,citecolor=blue,pdftitle={Title},pdfauthor={},pdfdisplaydoctitle=true,pdfduplex=DuplexFlipLongEdge]{hyperref}
\usepackage{braket}
\usepackage{float}
\usepackage{overpic}
\usepackage{gensymb}
\usepackage{mathrsfs}
\usepackage{tikz}
\usepackage{bm}
\usepackage{multirow}
\usepackage{amsfonts}
\usepackage{amsmath}
\usepackage{mathrsfs}
\usepackage{amssymb}
\usepackage{paralist}
\usepackage{indentfirst}
\usepackage{ulem}
\usepackage{soul}
\usepackage{cancel}
\usepackage{CJK}

\allowdisplaybreaks

\hypersetup{
           breaklinks=true,   
           colorlinks=true,   
           pdfusetitle=true,  
        }

\begin{document}

\title{Realization of a period-$3$ coplanar state in one-dimensional spin-orbit coupled optical lattice}

\author{Yida Chu}
\affiliation{Department of Physics and Chongqing Key Laboratory for Strongly Coupled Physics, Chongqing University, Chongqing 401331, China}
\affiliation{Beijing Computational Science Research Center, Beijing 100084, China}
\affiliation{Center of Modern Physics, Institute for Smart City of Chongqing University in Liyang, Liyang, 213300, China}

\author{Shijie Hu}
\email[]{shijiehu@csrc.ac.cn}
\affiliation{Beijing Computational Science Research Center, Beijing 100084, China}
\affiliation{Department of Physics, Beijing Normal University, Beijing, 100875, China}

\author{Tao Wang}
\email[]{tauwaang@cqu.edu.cn}
\affiliation{Department of Physics and Chongqing Key Laboratory for Strongly Coupled Physics, Chongqing University, Chongqing 401331, China}
\affiliation{Center of Modern Physics, Institute for Smart City of Chongqing University in Liyang, Liyang, 213300, China}
\affiliation{State Key Laboratory of Quantum Optics and Quantum Optics Devices, Shanxi University, Taiyuan, 030006, China}

\begin{abstract}
In ultracold atoms, achieving a period-$3$ structure poses a significant challenge.
In this work, we propose a three-sublattice spin-flop transition mechanism, differing from the two-sublattice counterpart used to explain the emergence of ferrimagnetic orders in higher dimensions.
Guided by this mechanism, we design a setup of alkaline-earth-metal atoms to create a spin-orbit coupled optical lattice, where we identify a triplet-fold degenerate $YX\bar{Y}$ state with a period-$3$ coplanar spin ordering within the deep Mott-insulating phase region of the ground-state phase diagram.
The $YX\bar{Y}$ state is protected by a finite gap, and its characteristic angle can be finely tuned by specific setup parameters.
Moreover, we use the Rabi spectroscopy technique to detect the $YX\bar{Y}$ state.
Our work not only shows the feasibility of achieving a period-$3$ structure \textit{via} the new mechanism but also suggests its potential applications for exploring other periodic structures in optical lattices.
\end{abstract}
\maketitle

Charge density waves (CDW) and spin density waves (SDW), known for their spatial modulation with two components $a$ and $b$, are fundamental in condensed matter physics~\cite{Dong_2008, Cho_2018}.
Introducing a third distinct component $c$ enables the emergence of chirality, resulting in a period-$3$ structure~\cite{Chepiga_2019}.
This concept is exemplified by the noncoplanar chiral CDW observed in three-dimensional materials like $1$T-TiSe$_2$~\cite{Ishioka_2010}, and the chiral SDW in the frustrated square lattice Hubbard model~\cite{Huang_2020}.
These phenomena interact with superconductivity in magic-angle-twisted bilayer graphene~\cite{Liu_2018},
exhibit nonlinear light coupling in the Weyl semimetal (TaSe$_4$)$_2$I~\cite{Gooth_2019},
and can lead to the formation of Chern insulators due to nontrivial Berry phases~\cite{Huang_2020}.
Understanding the chirality in period-$3$ structures is crucial for exploring the two-dimensional continuous commensurate melting transition~\cite{Huse_1982, Huse_1984, Bartelt_1987, Schreiner_1994}.
This understanding extends to models like the three-state chiral clock model~\cite{Selke_1982, Bartelt_1987, Dai_2017, Whitsitt_2018}, the three-state chiral Potts model~\cite{Selke_1982, Albertini_1989, Baxter_1989, McCoy_1990, Cardy_1993}, and parafermion physics~\cite{Fendley_2012,  Fendley_2014, Zhuang_2015}, coplanar phases observed in the spin-$1/2$ triangular-lattice Heisenberg antiferromagnet Ba$_3$CoSb$_2$O$_9$~\cite{Ito_2017}.
Moreover, a locally commensurate CDW with three-unit-cell periodicity has been observed in YBa$_2$Cu$_3$O$_y$~\cite{Vinograd_2021}, in contrast to the common period four in other hole-doped high-$T_c$ materials~\cite{Lee_2006}.

The creation of a period-$2$ superlattice, such as the ``$abab$" pattern in optical lattices of double wells, has been successfully achieved using four laser beams with different wavelengths~\cite{Sebby_Strabley_2006, F_lling_2007, Schreiber_2015}.
Similarly, the realization of a structure like ``$aabaab$" in optical lattices is relatively straightforward, which is accomplished by employing optical parametric oscillation (OPO) technique to generate a laser beam with a three-fold wavelength and then locking its phase to match that of the original laser beam.
However, creating a period-$3$ structure like ``$abcabc$" in optical lattices remains challenging, which requires more complicated OPO and phase-locking techniques.
Specifically, it involves generating a strong laser with a certain wavelength, a weaker laser with a two-fold wavelength, and an even weaker laser with a three-fold wave length, while ensuring that the initial phases of these three laser beams are precisely locked.
Alternatively, in an optical tweezer array, adding a spatial phase modulator to control the laser-atom coupling strength and phase at each site is necessary but challenging.
To date, this has never been experimentally achieved in one-dimensional ($1$D) optical lattice, particularly through mechanisms involving spontaneous symmetry breaking.
Additionally, the feasibility of realizing a complex SU($3$) Hubbard model with ultracold alkaline-metal and alkaline-earth-metal atoms has been investigated, with the potential to access the period-$3$ CDW phase through precise manipulation of interaction strengths~\cite{Honerkamp_2004, Gorshkov_2010, Taie_2012, Richaud_2022}.
Nevertheless, theoretical studies in advance suggest that period-$3$ coplanar CDW orders may emerge independently of chirality~\cite{Feng_2023, SU3_EPJB}.
In this work, we propose a three-sublattice spin-slope mechanism for achieving a three-period SDW in a spin-orbit coupled optical lattice of alkaline-earth-metal atoms and present a detection scheme.

\begin{figure}[!t]
\includegraphics[width=\linewidth]{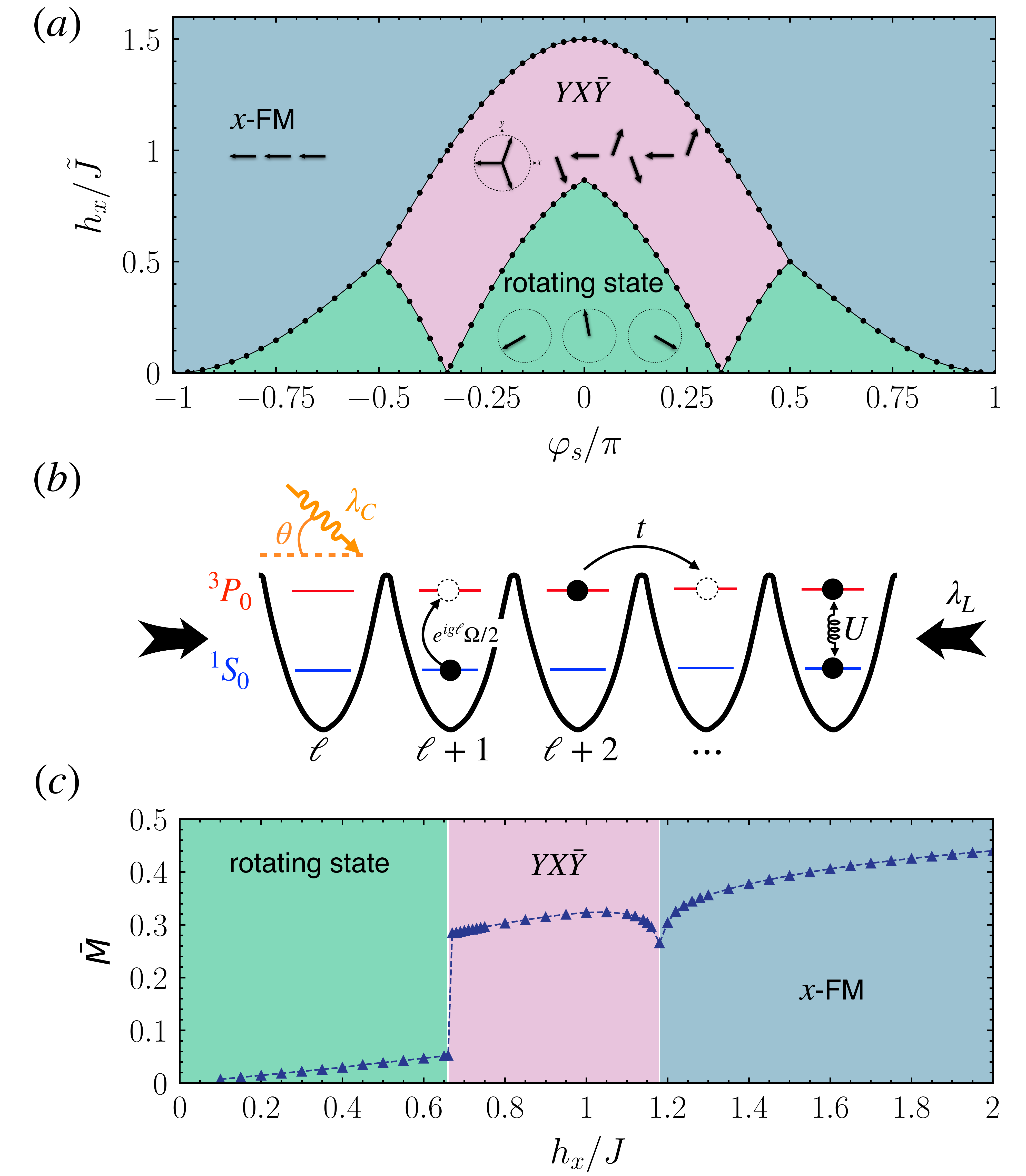}
\caption{(Color online)
(a) The mean-field phase diagram of model~\eqref{eq:MFHam} features three distinct states: a rotating state, a $YX\bar{Y}$ state, and a $x$-FM state.
(b) A schematic picture of the setup for using alkaline-earth-metal atoms to realize the model~\eqref{eq:Ham}~\cite{OLC_YJ_Nature, OLC_ARey_PRL, Yb_Fallani_PRL}.
In this setup, atoms are loaded into a $1$D optical lattice, where their internal states correspond to the long-lived clock states $^{1}S_0$ (blue) and $^{3}P_0$ (red).
The transition rate between these internal states at a site is controlled by the Rabi frequency $\Omega \ge 0$.
Atoms run within the lattice with a hopping amplitude $t > 0$, and strong repulsion with a strength of $U/t \gg 1$ occurs when two atoms occupy different internal states at the same site.
The angle $g=(\pi \lambda_\text{L} / \lambda_\text{C})\cos\theta$ is determined by the wavelengths $\lambda_\text{L}$ and $\lambda_\text{C}$ of the clock and the lattice lasers, and the angle $\theta$ between them. 
(c) The magnetization $\bar{M}$ calculated \textit{via} DMRG when applying an external magnetic field along the $x$-axis to the effective model~\eqref{eq:EffHam} with $D/J=1.5$ and $\Delta/J=-1.4$ under periodic boundary conditions.
A truncated bond dimension $m=512$ and a chain length $L=192$ are used.
\label{fig:fig1}
}
\end{figure}

\textit{Three-sublattice spin-flop transition}.
The spin-flop transition represents the abrupt reorientation of local magnetic moments driven by an external magnetic field, commonly observed in two-sublattice easy-axis antiferromagnets in both two and three dimensions~\cite{Neel_1936, Neel_1952}. 
Our findings indicate that this effect has practical implications in three-sublattice cases, particularly in the construction of period-$3$ coplanar spin order in one-dimensional ($1$D) systems.

Consider a chain of $L$ spinhalves.
We begin with a mean-field (MF) wave function $\ket{\psi_\text{MF}} = \otimes^L_{\ell=1} \ket{\psi_\ell}$, which describes coplanar spin ordering in the $xy$-plane.
At site-$\ell$, the spin wave function is given by $\ket{\psi_\ell} = (\ket{\uparrow_\ell} + e^{i \phi_\ell} \ket{\downarrow_\ell})/\sqrt{2}$, where $\phi_\ell \in [0,\, 2\pi)$ is a tunable angle, and $\uparrow_\ell$ and $\downarrow_\ell$ represent the spin-up and down along the $z$-axis, respectively.
In the presence of an external magnetic field along the $x$-axis, the energy per site is described by the functional
\begin{eqnarray}\label{eq:MFHam}
e \left(\{\phi_\ell\}\right) = \frac{1}{L} \sum^L_{\ell=1} \left[ \frac{\tilde{J}}{4} \cos \left(\phi_\ell - \phi_{\ell + 1} + \varphi_\text{s} \right) + \frac{h_x}{2} \cos \phi_\ell \right]\, ,
\end{eqnarray}
which depends on the angles $\{\phi_\ell\}$ and $\varphi_\text{s}$, given the coupling strength $\tilde{J} > 0$ between neighboring spin-halves and the magnitude of magnetic field $h_x \ge 0$.

In general, three cases can be identified:
(\textbf{i}) To minimize the energy per site $e$ when $h_x = 0$, it is advantageous for the angles $\{\phi_\ell\}$ to satisfy the condition $\phi_{\ell + 1} - \phi_\ell = \varphi_\text{s} - \pi$ (mod $2\pi$) at site $\ell$, as illustrated in Fig.~\ref{fig:fig1}(a).
When this condition is met, the system is in the ground state known as the \textit{rotating state}, with an average energy $e_\text{r} = -\tilde{J} / 4$.
In this state, there is no magnetic polarization along the $x$-axis to first-order perturbation.
(\textbf{ii}) Beyond a critical magnetic field magnitude $h^\text{sat}_x$, the system adopts a ground state where all spins align antiparallel along the $x$-axis, with $\phi_\ell = \pi$ for all sites.
The energy per site in this state is given by $e_\text{sat} = (\tilde{J} / 4) \cos \varphi_\text{s} - (h_x/2)$, and it is referred to as the \textit{$x$-FM state}.
(\textbf{iii}) In the intermediate region, a period-$3$ \textit{$YX\bar{Y}$ state} emerges when $L$ is a multiple of $3$.
This state is characterized by three variational angles: $\phi_{3n + 1} \equiv \phi_a$, $\phi_{3n + 2} \equiv \phi_b$, and $\phi_{3n + 3} \equiv \phi_c$ for the sublattice sites $a$, $b$, and $c$ within the $n$-th unit cell, as shown in Fig.~\ref{fig:fig1}(a).
Minimizing $e$ suggests a minimum
\begin{eqnarray}\label{eq:ep3}
e_{YX\bar{Y}} = \frac{\tilde{J}}{12} \left[\cos(2\varphi +\varphi_\text{s}) - 2\cos(\varphi - \varphi_\text{s})\right] + \frac{h_x}{6} (2\cos\varphi - 1)\, ,\nonumber
\end{eqnarray}
where we have $\phi_b = \pi$, $\phi_a = -\phi_c = \varphi$ with the characteristic angle $\varphi$ of the $YX\bar{Y}$ state.
Compared to the rotating state, the $YX\bar{Y}$ state has higher kinetic energy due to the disruption of the uniform angle gradient over a long distance.
When $h^{\phantom{\dag}}_x > h^{(\text{sf})}_x$, the gain in kinetic energy outweighs the potential energy loss from the magnetic field, thereby stabilizing the $YX\bar{Y}$ state.

By comparing $e_\text{r}$, $e_\text{sat}$, and $e_{YX\bar{Y}}$ across distinct states, we construct a MF phase diagram shown in Fig.~\ref{fig:fig1}(a).
When $\lvert \varphi_\text{s} \rvert > \pi / 3$, a first-order transition occurs from the rotating state to the $x$-FM state at $h^\text{sat}_x = (\tilde{J} / 2) (1 + \cos \varphi_\text{s})$, causing all spins to abruptly align along the $x$-axis.
In the region $\lvert \varphi_\text{s} \rvert < \pi / 3$, two successive transitions are observed as the magnetic field magnitude $h_x$ is tuned.
At the \textit{spin-flop transition} point $h^\text{sf}_x \ne 0$, the rotating state transitions to the $YX\bar{Y}$ state, which is guaranteed by $\cos \phi_a + \cos \phi_b + \cos \phi_c \ne 0$, or equivalently, $\varphi \ne \pi/3$.
Additionally, at another transition point $h^{\text{sat}}_x$, the $YX\bar{Y}$ state has a critical angle $\varphi_c \ne \pi$.

\textit{A setup of alkaline-earth-metal atoms}.
To induce the spin-flop transition and obtain the $YX\bar{Y}$ state, we propose loading fermionic alkaline-earth-metal atoms, such as $^{87}$Sr and $^{171}$Yb, into $1$D optical lattices (see Fig.~\ref{fig:fig1}(b)).
This setup allows us to simulate the spin-orbit coupling (SOC) by employing an additional clock laser beam~\cite{OLC_YJ_Nature, OLC_ARey_PRL, Yb_Fallani_PRL}.
In the setup, the motion of atoms is described by the Hamiltonian
\begin{equation}\label{eq:Ham}
\begin{aligned}
    H &=-t \sum^L_{\ell=1} \left( s_\ell^\dag s_{\ell+1}^{\phantom{\dag}} + p_\ell^\dag p_{\ell+1}^{\phantom{\dag}} + \textrm{H.c.} \right) + U \sum^L_{\ell=1} n^s_\ell n^p_\ell \\
    & + \frac{\Omega}{2} \sum^L_{\ell=1} \left( e^{i g \ell} p_\ell^\dag s_\ell^{\phantom{\dag}} + \textrm{H.c.} \right)\, ,
\end{aligned}
\end{equation}
where $s_\ell^{\phantom{\dag}}$ ($p_\ell^{\phantom{\dag}}$), $s_\ell^\dag$ ($p_\ell^\dag$) and $n^s_\ell=s_\ell^\dag s_\ell^{\phantom{\dag}}$ ($n^p_\ell=p_\ell^\dag p_\ell^{\phantom{\dag}}$) denote the annihilation, creation and particle number operators for the atom staying at the long-lived clock ground state $^{1}S_0$ (first-excited state $^{3}P_0$), respectively.
The third term is the Rabi term caused by the clock laser beam.
The wavelength of the clock laser is incommensurate with the lattice spacing, thus a site-dependent phase shift with an angle $g \ne 0$ is introduced.

\begin{figure*}[!t]
\includegraphics[width=\linewidth]{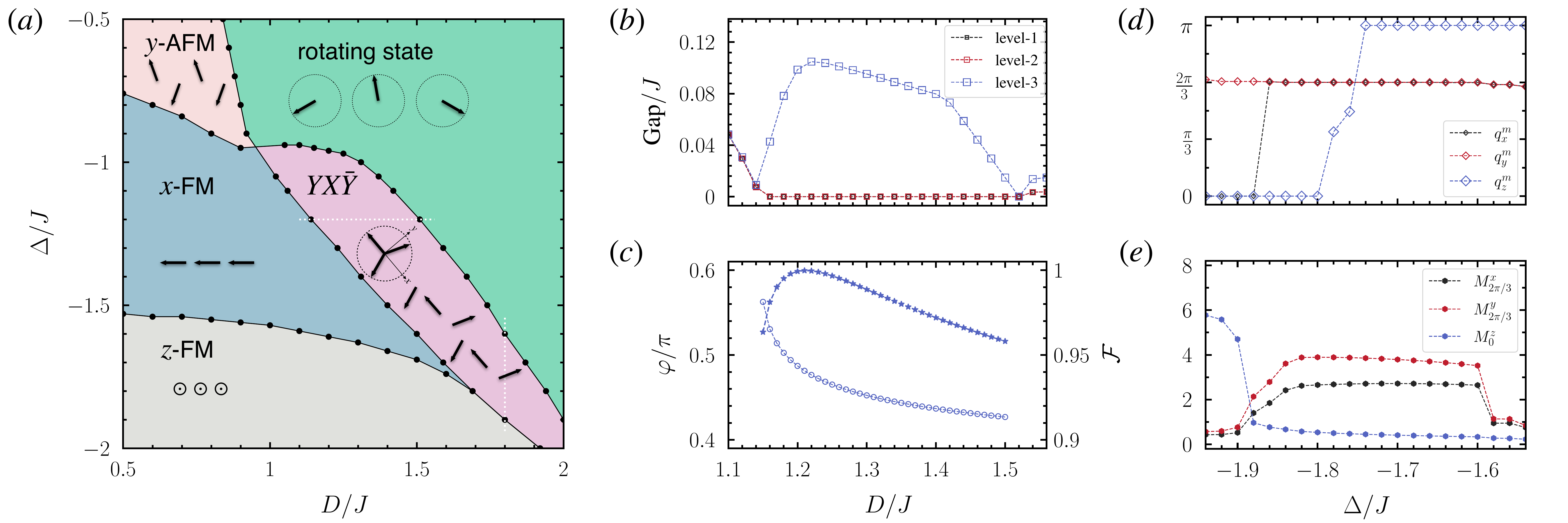}
\caption{(Color online)
(a) The groundstate phase diagram of the spin model~\eqref{eq:EffHam} at $h_x / J = 1$ that is determined by the DMRG calculations.
The negative-$\Delta$ region involves five states: $y$-AFM, $x$-FM, $z$-FM, rotating, and $YX\bar{Y}$.
(b) The gaps for the energy level-$1$, $2$, and $3$, (c) the characteristic angle $\varphi$ ($\circ$) and the fidelity $\mathcal{F}$ ($\star$) for the $YX\bar{Y}$ state as a function of $D/J$ at $\Delta / J=-1.2$.
(d) Peak momenta and (e) specific components of the structure factor as a function of $\Delta / J$ at $D / J=1.8$.
We use the periodic and open boundary conditions for (b, c) and (d, e), respectively, and a truncated bond dimension $m=512$ and $L=192$ for both.
\label{fig:fig2}}
\end{figure*}

At \textit{half filling}, the effective Hamiltonian in the deep Mott-insulating phase is given by~\cite{map_ARey_PRR}
\begin{equation}\label{eq:EffHam}
\begin{aligned}
    H_\text{eff} &= \frac{J}{4} \sum^L_{\ell=1} \left( \sigma^x_\ell \sigma^x_{\ell+1} + \sigma^y_\ell \sigma^y_{\ell+1} \right) + \frac{\Delta}{4} \sum^L_{\ell=1} \sigma^z_\ell \sigma^z_{\ell+1} \\
    &+ \frac{D}{4} \sum^L_{\ell=1} \left( \sigma^x_\ell \sigma^y_{\ell+1} - \sigma^y_\ell \sigma^x_{\ell+1} \right) + \frac{h_x}{2} \sum^L_{\ell=1} \sigma^x_\ell\, .
\end{aligned}
\end{equation}
Here, we use Pauli operators $\sigma^{x,y,z}_\ell$ in the spinor representation, with the bases $\ket{\uparrow_\ell} \equiv p^\dag_\ell \ket{0_\ell}$ and $\ket{\downarrow_\ell} \equiv e^{-i g \ell} s^\dag_\ell \ket{0_\ell}$, where $\ket{0_\ell}$ denotes the vacuum state.
In the Hamiltonian~\eqref{eq:EffHam}, $J$ gives the $XY$ coupling, $\Delta$ represents the $z$-axis anisotropy of the spin exchange, $D$ is the effective $z$-axis Dzyaloshinskii-Moriya (DM) interaction strength.
These parameters are determined by $t$, $U$, $g$, and $\Omega$ as follows:
\begin{eqnarray}\label{eq:SpinPara}
\begin{split}
J &= \frac{4t^2}{U} \frac{U^2 \cos g - \Omega^2 \cos^2\left(\frac{g}{2}\right)}{U^2-\Omega^2}\, ,\quad h_x = \Omega - \frac{4 t^2 \Omega \sin^2 \left(\frac{g}{2}\right)}{U^2-\Omega^2}\, ,\\
D &= \frac{2 t^2}{U} \frac{(\Omega^2 - 2 U^2) \sin g}{U^2-\Omega^2}\, ,\quad\quad\ \Delta = \frac{4t^2}{U} \frac{U^2 - \Omega^2 \cos^2 \left(\frac{g}{2}\right)}{U^2-\Omega^2}\, .
\end{split}\nonumber
\end{eqnarray}
This spin model~\eqref{eq:EffHam} has been extensively studied both analytically and numerically~\cite{Spin_Afflect_PRB, Spin_JHC_PRB}, primarily in the region where $D$ is small relative to $J$.
This is because, in real materials, the SOC is typically weak, with $D/J \sim 10^{-3}$~\cite{DM_D_JPCS, DM_M_PR}.
However, in the highly controllable setup for ultracold alkaline-earth-metal atoms, achieving a strong SOC is feasible, where $D$ can be comparable to $J$~\cite{SM}.
In the model~\eqref{eq:EffHam}, the first and third terms combine to effectively generate the parameter $\tilde{J} = \sqrt{J^2 + D^2}$ and $\varphi_\text{s} = \arctan (D/J)$ as defined in MF~\eqref{eq:MFHam}.

In the limit of dominantly-large $D$, the system is boosted by large $\tilde{J}$ and prefer the rotating state, which features a spiral configuration with an incommensurate rotation angle in the $xy$-plane~\cite{Spin_Afflect_PRB, Spin_JHC_PRB, Spin_Sun_PRB}.
{As the magnetic field magnitude $h_x$ can be freely tuned by the Rabi frequency $\Omega$, the ground state can enter the region for the $x$-FM state in the limit of large $h_x$.
At appropriate parameters $J = 1$, $D=1.5$ and $\Delta=-1.4$, we use the density matrix renormalization group (DMRG) method~\cite{DMRG_White_PRB, DMRG_Schollwock_RMP, DMRGMPS_Schollwock_AOP} to calculate the magnetic structure factor for the ground state $M_{\bm{q}} = \sqrt{(M^x_{q_x})^2 + (M^y_{q_y})^2 + (M^z_{q_z})^2}$.
Here, the $\alpha$ ($= x$, $y$, and $z$) component is given by
\begin{eqnarray}\label{eq:magnetization}
M^\alpha_{q_\alpha} = \sqrt{\frac{1}{4 L} \left\lvert \sum_{\ell, \ell^\prime = 1}^L e^{i q_\alpha (\ell - \ell^\prime)} \braket{\sigma_\ell^\alpha \sigma_{\ell^\prime}^\alpha} \right\rvert}
\end{eqnarray}
with $q_\alpha$ representing the $\alpha$ component of the magnetic vector ${\bm q} = (q_x,\, q_y,\, q_z)$.
At ${\bm q}^\text{m}$, the structure factor reaches its maximum value, referred to as the magnetization $\bar{M}$.
The $\bar{M}$ curve, shown in Fig.~\ref{fig:fig1}(c), reveals a spin-flop transition at $h^\text{sf}_x / J \approx 0.66$ and another transition at $h^\text{sat}_x / J \approx 1.18$.
In both the rotating and $x$-FM states, the peaks are situated at ${\bm q}^\text{m} = (0,\, 0,\, 0)$.
While in the $YX\bar{Y}$ state, ${\bm q}^\text{m}_x = {\bm q}^\text{m}_y = 2\pi/3$, providing a hallmark of a period-$3$ structure of spin orientations in the $xy$-plane.

Let us analyze the groundstate phase diagram for $h_x / J = 1$.
Figure~\ref{fig:fig2}(a) reveals that the presence of the DM term leads to the emergence of both the rotating state and the $YX\bar{Y}$ state across a broad range of parameters, unlike its absence in previous studies~\cite{XXZ_Dmitriev_JETP, XXZ_Dmitriev_PRB}.
Figure~\ref{fig:fig2}(b) illustrates the energy gap of the lowest-lying excited states, demonstrating that the $YX\bar{Y}$ state exhibits a triple-fold degeneracy, a consequence of spontaneous breaking of discrete translation symmetry.
This finite gap plays a crucial role in stabilizing the $YX\bar{Y}$ state against the external noise.
Compared to the rotating state, which lacks such a finite gap, the $YX\bar{Y}$ state may be more readily detectable.
Figure~\ref{fig:fig2}(c) also shows that the characteristic angle $\varphi$, defined as the spin angle marked in the cartoon of the $YX\bar{Y}$ state at the sublattices $a$ and $c$, exhibits sensitivity primarily to the strength of the DM interaction strength $D$, rather than to the anisotropy strength $\Delta$.
This sensitivity arises because $D$ directly influences the effective angle $\varphi_\text{s}$ in the MF scenario~\eqref{eq:MFHam}.
We also find that the MF wave function $\ket{\psi_\text{MF}}$ performs excellently in describing the $YX\bar{Y}$ state by evaluating its fidelity $\mathcal{F} = \vert \braket{\psi_\text{MF} \vert \psi} \vert$ with respect to the many-body wave function $\ket{\psi}$ obtained from the DMRG calculations.
As shown in Fig.~\ref{fig:fig2}(c), at $\Delta=1.8$, we can see that for all values of $D$, $\mathcal{F}$ consistently remains larger than $0.95$, reaching its maximum at $D \approx 1.21$, which approximates to $1$.

The $YX\bar{Y}$ state exhibits rich magnetic structures within the structure factor components $M^\alpha_{q_\alpha}$.
In Fig.~\ref{fig:fig2}(e), in the $YX\bar{Y}$ state region where $-1.88 \lesssim \Delta / J \lesssim -1.6$ and $D/J = 1.8$, the dominant contribution to the magnetization $\bar{M}$ comes from the components $M^x_{2\pi/3}$ and $M^y_{2\pi/3}$ in the $x$ and $y$ axes, respectively.
These two components have slight differences and correspond to intricate long-range correlations in the $xy$-plane.
Additionally, Figure~\ref{fig:fig2}(d) reveals that the peak $q^\text{m}_z$ of the component $M^z_{q_z}$ undergoes a rapid crossover from $\pi$ to $0$ as $\Delta$ decreases, near $\Delta / J \approx -1.74$.
In the rotating state, the emerging short-range antiferromagnetic (AFM) ordering along the $z$-axis, with $q^\text{m}_z = \pi$, competes with the irrelevant $z$-axis anisotropic terms with $\Delta < 0$ in the model~\eqref{eq:EffHam}, destabilizing the ground state by increasing its energy.
This effect facilitates a spin-flop transition from the rotating state and the $YX\bar{Y}$ state when a magnetic field is applied along the $x$-axis.
In contrast, in the positive-$\Delta$ region, this effect is absent, preventing the appearance of the $YX\bar{Y}$ state.
As $\Delta$ decreases further, the system transitions to $z$-FM with $q^\text{m}_z = 0$, where spins align parallelly in the $z$-axis.

When $\Delta/J \gtrsim -1$, applying $h_x$ causes a two-sublattice spin-flop transition from the rotating state to the $y$-AFM state, in which spins on adjacent sites align parallel along the $x$-axis and antiparallel along the $y$-axis, as shown in the upper-left corner of the phase diagram in Fig.~\ref{fig:fig2}(a).
In higher dimensions, particularly in bipartite lattices such as the square lattice, interactions along other directions often stabilize a period-$2$ structure.
However, in $1$D, a period-$3$ $YX\bar{Y}$ state can survive over a broad region and is more stable than period-$2$ $y$-AFM when $\Delta / J \lesssim -1$.

\begin{figure}[!t]
\includegraphics[width=\linewidth]{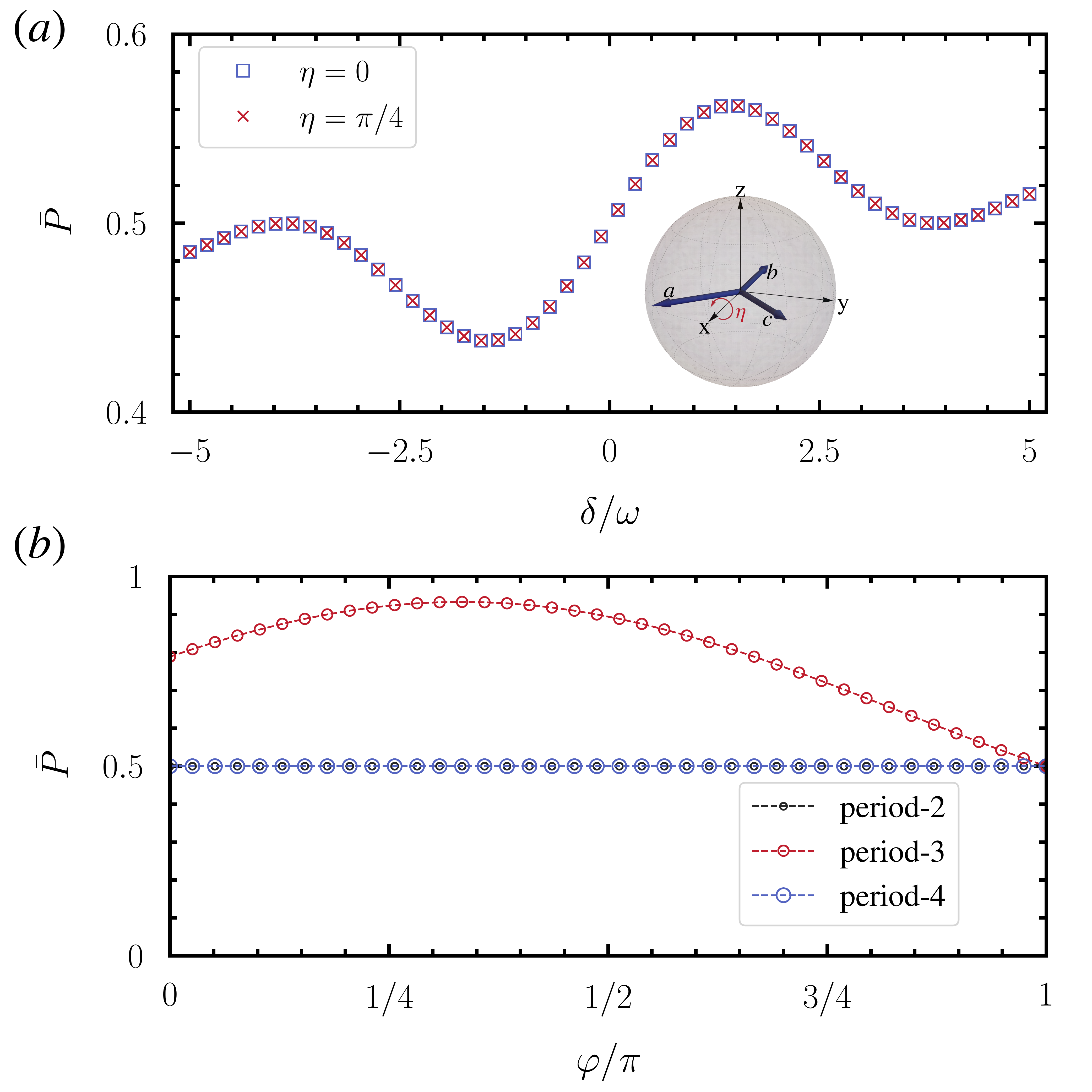}
\caption{(Color online)
(a) Rabi spectrum $\bar{P}$ as a function of the detuning $\delta$ of the driving clock laser, after rotating all spins by angles $\eta=0$ ({\color{blue}$\square$}) and $\pi/4$ ({\color{red}$\times$}) about the $x$-axis.
Inset: before measurement, another clock laser is applied to rotate the spins, with the black vectors representing the spin orientations at three sublattice sites a, b, and c, in the $YX\bar{Y}$ state.
(b) Rabi spectrum $\bar{P}$ as a function of the characteristic angle $\varphi$ for period-$2$ ({\color{black}$\circ$}), period-$3$ ({\color{red}$\circ$}), and period-$4$ ({\color{blue}$\circ$}) spin structures, respectively.
For (a), we use $g^\prime = g$, while for (b), $g^\prime = g + 2\pi/3$. For (a), we set the frequency $\omega$ of the driving clock laser as the energy unit.
\label{fig:fig3}
}
\end{figure}

\textit{Rabi spectroscopy}.
Next, we propose using Rabi spectroscopy technique~\cite{PhysRevA.80.052703} to identify the $YX\bar{Y}$ state from other distinct states in the ground-state phase diagram shown in Fig.~\ref{fig:fig1}.
The detailed scheme involves two rounds of measurement to provide the probability $\bar{P}$ for atoms in the $^{3}P_0$ ($\uparrow$) level after time evolution under a $\pi/2$-pulse of the driving clock laser, averaged over all sites, named the ``Rabi spectrum".

In the first round, we begin by rotating all spins by an angle $\eta$ about the $x$-axis using a zero-detuned clock laser pulse, as illustrated in the inset of Fig.~\ref{fig:fig3}(a).
We then set the same phase shift $g^\prime = g$ in the spectroscopy technique and measure the Rabi spectrum $\bar{P}$ as a function of the detuning $\delta$ of the driving clock laser.
For the $YX\bar{Y}$, $x$-FM, $y$-AFM states, where spin orientations at sublattice sites are symmetric about the $x$-axis, the Rabi spectra are identical for all values of $\eta$~\cite{SM}, exemplified with $\eta=0$ and $\pi/4$ in Fig.~\ref{fig:fig3}(a).
Notably, the detection round also helps us exclude the $z$-FM state, which show distinct Rabi spectra for different $\eta$~\cite{SM}.
In the second round, we use the driving clock laser to create a light field that interacts with the spins, inducing an effective period-$3$ modulation relative to the site-dependent phase shift in the model~\eqref{eq:Ham}.
This is achieved by setting a specific angle $\theta^\prime$ between the clock laser and the lattice laser, yielding an angle of $g^\prime = g + 2\pi/3$.
Next, we measure the Rabi spectrum $\bar{P}$ as a function of the characteristic angle $\varphi$.
As shown in Fig.~\ref{fig:fig3}(b), only the ground state with a period-$3$ spin structure shows a clear dependence of $\varphi$, with $\bar{P}$ varying with $\varphi$ or $D / J$ in practice.
For other states with periods different from $3$, $\bar{P}$ remains constant at $1/2$, regardless of $\varphi$.
By analyzing this Rabi spectrum, we can identify the ground state with a period-$3$ spin structure, such as the $YX\bar{Y}$ state.
After these two rounds of measurements, we can identify the $YX\bar{Y}$ state, characterized by spin orientations symmetrically distributed about the $x$-axis with a period-$3$ spatial profile.
Notably, although the rotating state has strong quantum fluctuations in $1$D, we can still rule it out using the above two-round meansurement~\cite{SM}.

\textit{Summary and discussion}.
We propose a three-sublattice spin-flop mechanism for achieving a period-$3$ $YX\bar{Y}$ coplanar state in a $1$D system of alkaline-earth-metal atoms.
To effectively detect this state, we recommend using two rounds of Rabi spectroscopy measurements, which offer sharper signals compared to the traditional two-sublattice mechanism used in higher-dimensional ferrimagnets.
This innovative mechanism can also be extended to create more complex periodic structures in various artificial setups, including optical lattices, arrays of Rydberg atoms, and beyond.

We are grateful to Xue-Feng Zhang, Zi-Jian Xiong, Wei Su, Shi-Ju Ran, Wei Wang, and Yan-Hua Zhou for fruitful discussions. National Science Foundation of China under Grant No.~12274045, No.~12347101, No.~12174020, MOST~2022YFA1402700, and NSAF ~U1930402, the Program of State Key Laboratory of Quantum Optics and Quantum Optics Devices (No.~KF202211) and Fundamental Research Funds for the Central Universities Grant No.~2023CDJZYJH-048.
Computational resources from Tianhe-2JK at the Beijing Computational Science Research Center and Quantum Many-body-{\rm I} cluster at SPA, Shanghai Jiaotong University are also highly appreciated.
\bibliographystyle{apsrev4-1}
\bibliography{reference}
\end{document}